\documentclass[fleqn,usenatbib]{mnras}
\usepackage{newtxtext,newtxmath}
\usepackage[T1]{fontenc}
\usepackage{epsfig}
\usepackage{mathtools}
\usepackage{cuted}

\title[Surface roughness and lunar water desorption]{Implications of surface roughness in models of water desorption on the Moon}

\author[Bj\"{o}rn J. R. Davidsson \& Sona Hosseini]{
Bj\"{o}rn J. R. Davidsson,$^{1}$\thanks{E-mail: bjorn.davidsson@jpl.nasa.gov}
Sona Hosseini,$^{2}$
\\
$^{1}$Jet Propulsion Laboratory, California Institute of Technology,  M/S 183--401, 4800 Oak Grove Drive, Pasadena, CA 91109, USA\\
$^{2}$Jet Propulsion Laboratory, California Institute of Technology,  M/S 183--601, 4800 Oak Grove Drive, Pasadena, CA 91109, USA
}

\date{Accepted 2021 May 5. Received 2021 May 5; in original form 2021 February 9.}

\pubyear{2021}

\begin{document}
\label{firstpage}
\pagerange{\pageref{firstpage}--\pageref{lastpage}}
\maketitle

\begin{abstract}
The observed presence of water molecules in the dayside lunar regolith was an unexpected discovery and 
remains poorly understood. Standard thermophysical models predict temperatures that are too high for adsorbed 
water to be stable. We propose that this problem can be caused by the assumption of locally flat surfaces that is 
common in such models. Here we apply a model that explicitly considers surface roughness, and accounts for 
solar illumination, shadows cast by topography, self--heating, thermal reradiation, and heat conduction. We couple 
the thermophysical model to a model of first--order desorption of lunar surface water and demonstrate that surface 
roughness substantially increases the capability of the Moon to retain water on its sunlit hemisphere at any latitude, and within $45^{\circ}$ of the poles, 
at any time of the lunar day. Hence, we show that lunar surface roughness has a strong influence on lunar water adsorption 
and desorption. Therefore, it is of critical importance to take account of surface roughness to get an accurate picture 
of the amount of water on the Moon's surface and in its exosphere.

\end{abstract}


\section{Introduction} \label{sec_intro}

The first strong indications of the existence of water ice on the Moon came from the detection of hydrogen deposits at 
the lunar poles through neutron spectrometry by the Lunar Prospector spacecraft \citep{feldmanetal98},  
and from bistatic radar experiments of the same regions carried out by the Clementine spacecraft \citep{spudisetal98}. 
The hydrogen host responsible for the anomalous radar echoes was confirmed to be water ice when $\mathrm{H_2O}$ 
emission from the ejecta plume created by the planned impact of a Centaur rocket stage into the permanently 
shadowed south pole crater Cabeus was detected and measured by the ultraviolet, visual, and near--infrared 
spectrometers on the LCROSS spacecraft \citep{colapreteetal10}. At the time of Lunar Prospector and Clementine,  
lunar water had not been detected in the exosphere \citep{stern99}, and water ice was expected to be confined to the 
poles. Tentative signals were registered at mass 18 ($\mathrm{H_2O}$) and 17 (OH or $\mathrm{NH_2}$) by the 
Apollo~17 LACE mass spectrometer at the lunar surface before dawn \citep{hoffmanandhodges75}. However, 
these signals could have been the result of outgassing from either the instrument itself or the nearby lunar module, 
rather than the result of indigenous water. The detection of absorption features due to small amounts of $\mathrm{H_2O}$ and/or 
OH on the surface of the sunlit lunar hemisphere indicated the lunar water was not confined to the 
polar regions; as shown by the Cassini \citep{clark09}, Deep Impact \citep{sunshineetal09}, and 
Chandrayaan--1 \citep{pietersetal09} spacecraft. The Deep Impact observations demonstrated a clear 
anticorrelation between the $\mathrm{H_2O/OH}$ abundance and illumination levels, and thus an anticorrelation between the $\mathrm{H_2O/OH}$ 
abundance and surface temperature. The water surface coverage decreased from dawn to noon, then increased 
again from noon to dusk and it generally increased with distance from the equator \citep{sunshineetal09}. 
Observations also showed that the felsic highlands retained water more efficiently than the mafic Mare 
\citep{sunshineetal09, pietersetal09}, a phenomenon that also is seen in desorption experiments on lunar samples \citep{postonetal15}. 
The presence of $\mathrm{H_2O/OH}$ on the lunar dayside was confirmed and further studied with the LAMP instrument 
on the Lunar Reconnaissance Orbiter \citep{hendrixetal19}. Importantly, these observations showed that the dayside $\mathrm{H_2O/OH}$ remained 
during lunar passages through Earth's magnetotail, showing that it is not promptly produced by the solar wind but  adsorbed from the exosphere. 
These observations demonstrate  that forenoon injection of $\mathrm{H_2O/OH}$ from the surface into the exosphere and afternoon freeze--out of exospheric $\mathrm{H_2O/OH}$ 
onto the surface are complex phenomena and depend on the properties of the regolith, such as mineralogy, and 
the degree of surface roughness. Observations by \citet{honniballetal20} made at $6\,\mathrm{\mu m}$ 
show that molecular water (i.e., $\mathrm{H_2O}$) definitively is present on the lunar dayside.

The first direct detection of exospheric lunar water was made with the CHACE mass spectrometer onboard the Moon 
Impact Probe component of the Chandrayaan--1 spacecraft \citep{sridharanetal10, sridharanetal10b}. 
This short--duration experiment was followed by long--term \emph{in situ} mass spectrometric time--series measurements by 
the LADEE spacecraft that established a temporal correlation between spikes in the water abundance with the annual 
meteor showers \citep{bennaetal15, hurleyetal18}. Attempts to determine the steady--state abundance 
of exospheric water between meteorite--induced spikes with the HST \citep{sternetal97} and with the Chang'e--3 spacecraft 
\citep{wangetal15} were not successful as only upper limits on the column densities were obtained. In parallel to these observational efforts, 
the identification of indigenous water in lunar rock samples \citep[e.~g.,][]{saaletal08} contributed to a diversified view of the origin and history of lunar volatiles.

Substantial efforts have been made to model the structure and evolution of the lunar exosphere, typically through 
Monte Carlo models that track large numbers of individual molecular test particles as they are released from the surface, 
travel through the exosphere on ballistic trajectories, interact with the surface during bouncing or temporary residence, 
and eventually experience permanent cold entrapment or escape lunar gravity \citep[e.g.,][]{leblancandchaufray11, tenishevetal13, schorghofer14, schorghofer15, gravaetal15, schorghoferetal17}. 
In all these models, it is critical to know the surface temperature at each point on the lunar 
surface in order to assign appropriate molecular velocities during release or bouncing and to calculate molecular residence times 
(i.e., the amount of time a molecule is expected to remain on the surface at a given temperature before entering the exosphere). 
Typically, the treatment of lunar surface temperatures in these models falls into three categories where the temperature is 
obtained by: 1) usage of analytical formulae in which  the temperature is a simple function of latitude and local time \citep{leblancandchaufray11, tenishevetal13}; 
2) solving a one--dimensional energy conservation equation with a boundary condition that balances 
solar illumination, thermal reradiation, and heat conduction, assuming a locally flat surface \citep[e.g.][]{schorghofer14, schorghofer15, schorghoferetal17}; 
or 3) applying a temperature map derived from spacecraft observations \citep[e.g.][]{gravaetal15, schorghoferetal17}. Common to these models is that the locally applied temperature is an 
average that does not capture the significant small--scale temperature variability (amounting to $\Delta T\stackrel{<}{_{\sim}} 200\,\mathrm{K}$ on the size scale of a 
metre, see Sec.~\ref{sec_results}) that arise on real terrains due to the presence of local shadowed cold spots and 
self--heated hot spots caused by roughness. Individual molecules that interact with a rough terrain may end up in a cold spot and endure 
a residence time that is orders of magnitudes longer than calculated for the local average temperature, or it may bounce on a surface that 
is substantially hotter than the average and leave with unusually high velocity. Temperature maps based on spacecraft observations do not 
relax the temperature dispersion problem because they do not resolve the sub--metre--scale temperature fluctuations relevant for 
molecule--surface interactions. Furthermore, analytical functions of surface temperature fitted to spacecraft 
observations \citep[e.g.][]{hurleyetal15} can result in smoothing out the original data even more.

The paper by \citet{premetall18} aims at introducing a more realistic surface temperature treatment in exospheric 
modelling by accounting for surface roughness, although still in a simplified manner. In this model, the surface is represented 
by a large number of facets with individual tilts that collectively yield a certain level of roughness quantified by a mean slope angle. 
Because heat conduction is neglected, all facets are assumed to be in immediate radiative equilibrium with its surroundings, 
which means that zero incidence radiation flux would correspond to zero temperature. To avoid this issue, a realistic but arbitrarily 
chosen threshold temperature of $T\geq 130\,\mathrm{K}$ is applied. Mutual shadowing and self-heating are not calculated 
self--consistently for the assembly of facets. Instead, at a given solar incidence angle, a probability function is applied to determine 
the fraction of facets that should be considered shadowed. A fixed level of diffuse radiation is applied to all facets to account for 
self--heating (set so that the combined thermal energy loss of all facets should match the observed emission as closely as possible). 
It is difficult to assess the influence of assumptions like these on the calculated properties of the exosphere.  

\citet{hayneetal20} present a numerical model that evaluates the illumination conditions on rough terrains, including 
shadowing and self--heating without simplifying assumptions, and calculates the temperatures for these terrains, considering thermally 
conductive regolith. They applied the model in the context of estimating the percentage of the lunar surface that constitute permanent 
shadows. They find that 10--$20\%$ of all lunar cold trap area is in the form of small ($\geq 1\,\mathrm{cm}$) permanent shadows. 
The study by \citet{hayneetal20} shows the importance of roughness, and the value of applying appropriate models for 
understanding the behaviour of water on the Moon.

Another source of uncertainty in exospheric models is the evaluation of molecular residence time. The expressions used by several authors 
\citep[e.g.][]{leblancandchaufray11, premetall18} essentially quantify the time needed for removing a monolayer 
of molecules, albeit at a rate that applies for sublimation from multi--layer ice, which is much higher. Recognising that desorption 
of a monolayer of, e.g., water molecules from a rock substrate requires substantially longer time than the sublimation of a monolayer 
water molecules from a water ice substrate because of the somewhat different binding energies in the two cases, 
\citet{schorghofer14} increases the residence time by a factor 400.

The intention of this paper is to apply a realistic and physically self--consistent thermophysical model that accounts for 
solar illumination, thermal reradiation, shadowing, self--heating, and heat conduction of rough terrains without arbitrary 
assumptions that violate energy conservation. Furthermore, we attempt to deal with molecular residence time in a realistic 
manner, by numerically solving the Polanyi--Wigner equation in first--order mode that is appropriate for desorption of 
(sub)monolayers of volatiles. Our target is primarily to illustrate the difference between idealised flat terrains and realistic,
 rough terrains as it pertains to water molecule desorption behaviour in lunar conditions as a function of time after sunrise.  
According to our rough terrain model,  the existence and stability of water on specific places on the lunar surface are significantly 
more likely than previously supposed. This model should be seen as the first step in a long--term goal of building a lunar exosphere 
model that advances the state--of--the--art treatment of exosphere--surface interaction. This paper is structured as follows: the 
thermophysical model is described in Sec.~\ref{sec_model}; the desorption model is described in Sec.~\ref{sec_desorption}; 
our results showing that water can exist for an extended time on the lunar dayside are presented in Sec.~\ref{sec_results}; in 
Sec.~\ref{sec_discussion} we discuss our results and future measurements that can be used to advance the state of knowledge 
on water adsorption and desorption in lunar regolith; and a summary of conclusions is found in Sec.~\ref{sec_conclusions}.

\section{The thermophysical model} \label{sec_model}

In order to study the desorption of water molecules from the lunar surface, it is necessary to develop a description of how the 
surface temperature changes during lunar rotation. This is provided by a numerical thermophysical model that realistically accounts 
for the significant physical processes at work on the rough lunar surface: direct solar illumination, shadows cast by topography, 
visual and infrared self--heating (i.e., diffuse illumination emanating from other parts of the surface), thermal reradiation, and 
heat conduction. We here apply the thermophysical model developed by \citet{davidssonandrickman14}. This model has been 
used in order to determine the thermal inertia of Comet 9P/Tempel~1 by reproducing near--infrared emission measurements obtained 
by the Deep Impact spacecraft \citep{davidssonetal13}, to investigate the dependence of thermal emission properties on 
various topographic models of surface roughness \citep{davidssonetal15}, and to demonstrate that the neck on the bilobed 
Comet 67P/Churyumov--Gerasimenko is not likely to have developed by erosion \citep{sierksetal15}. \citet{davidssonetal15} 
also demonstrated the capacity of the model to reproduce the measured thermal infrared phase function of Asteroid (1) Ceres, and 
showed that a random Gaussian description of surface roughness more readily reproduced the incidence angle dependence of lunar 
thermal emission near zero emergence angle measured by the Diviner Radiometer on the Lunar Reconnaissance Orbiter, than other 
roughness models.

\begin{figure}
\begin{center}
     \scalebox{0.55}{\includegraphics{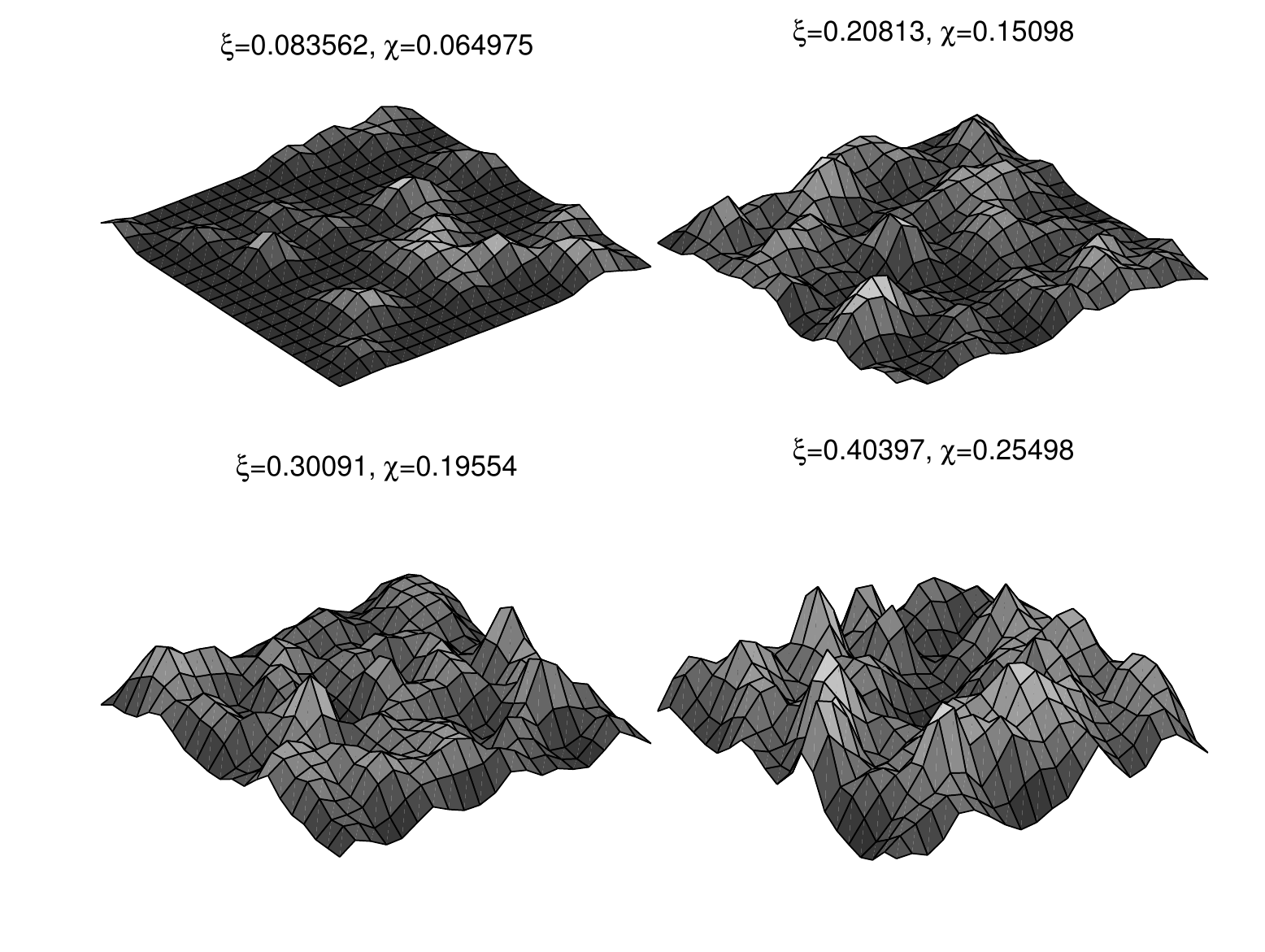}}
    \caption{The random Gaussian surface topographies used in our thermophysical modelling. The upper left surface has $\bar{\theta}=13.6^{\circ}$ and $s_{\rm rms}=22.5^{\circ}$, the upper right surface has $\bar{\theta}=23.6^{\circ}$ and $s_{\rm rms}=35.5^{\circ}$, 
the lower left surface has $\bar{\theta}=30.9^{\circ}$ and $s_{\rm rms}=42.9^{\circ}$, the lower right surface has 
$\bar{\theta}=38.6^{\circ}$ and $s_{\rm rms}=50.0^{\circ}$. These random models generally account for the variety of the 
small--scale variations on the surface of the Moon. Reprinted from Icarus, 243, B. J. R. Davidsson and H. Rickman, 
Surface roughness and three--dimensional heat conduction in thermophysical models, 58--77, Copyright (2014), with permission from Elsevier. 
This figure reproduces Fig.~10 of \protect\citet{davidssonandrickman14}.}
     \label{fig_terrains}
\end{center}
\end{figure}

We here briefly summarise the model of \citet{davidssonandrickman14}, for more details see that paper and the ones mentioned above. 
Surface roughness is modelled explicitly by generating a surface element that contains topographic detail, realised by a collection of 
triangular facets forming a continuous surface. The particular random Gaussian surface elements used in this work are shown in 
Fig.~\ref{fig_terrains}. The degree of surface roughness is measured in several different ways, e.g., by the Hapke mean slope angle 
$\bar{\theta}$ \citep{hapke84}, the RMS mean slope angle $s_{\rm rms}$ \citep{spencer90}, the small scale roughness 
parameter $\xi$, and the small scale self heating parameter $\chi$ \citep{lagerros97, lagerros98}. Numerical values in all these systems are provided in Fig.~\ref{fig_terrains}.

Measurements of the degree of roughness on the Moon has been done directly in images, as well as indirectly through model fitting of 
visual reflectance or thermal emission spectra. For example, \citet{helfensteinandshepard99} measured the roughness in 11 stereoimages of regolith 
acquired at the Apollo~11, 12, and 14 landing sites and found $12^{\circ}\leq s_{\rm rms}\leq 35^{\circ}$. In his analysis of lunar thermal emission 
spectroscopy, \citet{spencer90} obtained $s_{\rm rms}\approx 40^{\circ}$, while \citet{rozitisandgreen11} found $s_{\rm rms}=31.5^{\circ}$--$33^{\circ}$ 
in their study. \citet{bandfieldetal15} found that Lunar Reconnaissance Orbiter Diviner Radiometer data was best fitted if $20^{\circ}\leq s_{\rm rms}\leq 35^{\circ}$. 
The $s_{\rm rms}$--values for our first three synthetic terrains ($22.5^{\circ}$, $35.5^{\circ}$, $38.6^{\circ}$) therefore represent typical lunar terrain, 
while the $s_{\rm rms}=50^{\circ}$ case represents unusually rough regions.

The model of \citet{davidssonandrickman14} used for this study allows for three--dimensional heat conduction on rough 
terrains. Here, we consider heat conduction along surface normals only, and ignore lateral heat conduction. This simplification 
is justified as long as the facets on the surface element are thought of as being larger than the lunar diurnal skin depth, which 
is $L=\sqrt{P\kappa/2\pi c\rho}\approx 0.02\,\mathrm{m}$ (for model parameters and constants, see Tables~\ref{tab1} and \ref{tab2}). 
For each surface facet $j$ the model solves the energy conservation equation,
\begin{equation} \label{eq:01}
\rho c\frac{\partial T_j(x,t)}{\partial t}=\kappa\frac{\partial^2 T_j(x,t)}{\partial x_j^2}.
\end{equation}
The surface boundary condition of Eq.~(\ref{eq:01}) is given by

\begin{equation} \label{eq:02}
\begin{array}{c}
\displaystyle  \frac{S_{\odot}v_{j\odot}(1-A)\mu_j(t)}{r_{\rm h}^2}=\varepsilon\sigma T_j(0,t)^4-\kappa\frac{dT}{dx_j}\Big|_{x_j=0}\\
\\
\displaystyle - (1-A)\sum_{k\not=j}\mathcal{F}_{jk}\left(\frac{S_{\odot}v_{k\odot}A\mu_k(t)}{r_{\rm h}^2}+\varepsilon\sigma T_k(0,t)^4 \right).
\end{array}
\end{equation}
%

From left to right, the terms in Eq.~(\ref{eq:02}) denote: the absorbed flux from direct solar illumination; thermally emitted 
infrared radiation; heat conducted from the surface to the interior or vice versa depending on the sign of the temperature gradient;
 and self--heating in the form of scattered optical and thermally emitted radiation from other facets. 
We have here assumed that the absorptivity is the same both at optical and infrared wavelengths, which is not 
strictly correct, except for spectrally grey bodies. It is indeed $1-A$ in the visual but is $\varepsilon\not=1-A$ in the infrared. By absorbing 
$88\%$ of the available thermal self--heating radiation instead of $95\%$ (see Table~\ref{tab2}) we slightly underestimate the total self--heating. We 
expect this inconsistency to have a minor effect on the results. The cosine of the 
solar incidence angle of facet $j$, $\mu_j$, is calculated in accordance with the lunar spin pole in the equatorial system 
$\{\alpha_{\rm L},\,\delta_{\rm L}\}$, the latitude of the region on the lunar surface under consideration $l_{\rm L}$, the 
orientation of the facet in question, and the local time. Note, that an evaluation of whether the Sun is fully visible from a 
given facet $j$ ($v_{j\odot}=1$) or if the facet is partially (rounded to $v_{j\odot}=1/3$ or $v_{j\odot}=2/3$) or fully 
shadowed by topography ($v_{j\odot}=0$) is made at each time step. The view factors \citep[e.g.][]{ozisik85} 
\begin{equation} \label{eq:03}
\mathcal{F}_{jk}=\frac{v_{jk}s_k\cos\beta_j\cos\beta_k}{\pi r^2}
\end{equation}
determine what fraction of the flux scattered by facet $k$ in the visual, and emitted by it in the infrared, that reaches facet $j$. 
Here, $v_{jk}=1$ if facets $j$ and $k$ have unobstructed views of each other but $v_{jk}=0$ if their line of sight is interrupted 
by topography. Here, $s_k$ is the surface area of facet $k$, the angles between the surface normals of $j$ and $k$, and their 
interception line are $\beta_j$ and $\beta_k$, respectively, and $r$ is the distance between the two facets. The model of 
\citet{davidssonandrickman14} can account for multiple scattering, but we here apply the single--scattering formulation 
(Eq.~\ref{eq:02}) because of the relatively low lunar albedo. The boundary condition of Eq.~(\ref{eq:01}) at the bottom of the 
calculational domain (typically placed $\sim 10L$ below the surface) is a vanishing temperature gradient;
\begin{equation} \label{eq:04}
\frac{\partial T_j}{\partial x_j}\Big|_{x=x_{\rm max}}=0.
\end{equation}

\begin{table}
\begin{center}
\begin{tabular}{||l|l|l||}
\hline
\hline
Symbol & Description & Unit\\
\hline
$\mathcal{F}_{jk}$ & View factor & Dimensionless\\ 
$f$ & Fractional area coverage of $\mathrm{H_2O}$ & Dimensionless\\
$l_{\rm L}$ & Lunar latitude & $^{\circ}$\\ 
$L$ & Diurnal skin depth & $\mathrm{m}$\\
$r$ & Distance between facets & $\mathrm{m}$\\
$s_k$ & Facet area & $\mathrm{m^2}$\\
$s_{\rm rms}$ & RMS mean slope angle & $^{\circ}$\\ 
$T$ & Temperature & $\mathrm{K}$\\
$t$ & Time & $\mathrm{s}$\\
$v_{j\odot}$ & Facet/Sun visibility parameter & Dimensionless\\
$v_{jk}$ & Facet/facet visibility parameters & Dimensionless\\
$x$ & Depth coordinate & $\mathrm{m}$\\
$x_{\rm max}$ & Maximum depth & $\mathrm{m}$\\
$\beta$ & Angle: facet normal to direction  & $^{\circ}$\\ 
 & of other facet &\\
$\bar{\theta}$ & Hapke mean slope angle  & $^{\circ}$\\ 
$\mu$ & Cosine of solar incidence angle & Dimensionless\\
$\xi$ & Small scale roughness parameter & Dimensionless\\
$\chi$ & Small scale self heating parameter & Dimensionless\\
\hline 
\hline
\end{tabular}
\caption{List of parameters used in the thermophysical and desorption models in this paper.}
\label{tab1}
\end{center}
\end{table}

The surfaces considered in this work consist of 800 facets. We thus solve 800 differential equations (Eq.~\ref{eq:01}) coupled 
through their surface boundary conditions, Eq.~(\ref{eq:02}). Calculations are made for all four terrains shown in Fig.~\ref{fig_terrains} 
at three specific latitudes, $l_{\rm L}=0^{\circ}$ (the equator), $l_{\rm L}=45^{\circ}\,\mathrm{N}$, and $l_{\rm L}=70^{\circ}\,\mathrm{N}$. 
We also consider the $\xi\approx 0.2$ surface at $l_{\rm L}=85^{\circ}\,\mathrm{N}$.

\begin{table*}
\begin{center} {\bf } \end{center}
\begin{center}
\begin{tabular}{||l|l|r|l||}
\hline
\hline
Symbol & Description & Value & Unit\\
\hline
$A$ & Bond albedo & 0.12 & Dimensionless\\
$c$ & Specific heat capacity & $720$ & $\mathrm{J\,kg^{-1}\,K^{-1}}$\\
$\Delta E_{\rm des}$ & Activation energy & $8\cdot 10^{-20}$ & $\mathrm{J}$\\
$k_{\rm B}$ & Boltzmann constant & $1.3806\cdot 10^{-23}$ & $\mathrm{m^2\,kg\,s^{-2}\,K^{-1}}$\\
$n$ & Order of desorption & 1 & Dimensionless\\
$P$ & Synodic rotation period & $708.4$ & $\mathrm{h}$\\
$r_{\rm h}$ & Heliocentric distance & 1 & $\mathrm{AU}$\\
$S_{\odot}$ & Solar constant & $1367$ & $\mathrm{J\,m^{-2}\,s^{-1}}$\\
$\alpha_{\rm L}$ & Lunar spin pole right ascension & $266.8577$ & $^{\circ}$\\
$\delta_{\rm L}$ & Lunar spin pole declination & $65.6411$ & $^{\circ}$\\
$\varepsilon$ & Emissivity & $0.95$ & Dimensionless\\
$\kappa$ & Heat conductivity & $7.4\cdot 10^{-4}$ & $\mathrm{W\,m^{-1}\,K^{-1}}$\\
$\nu$ & Frequency factor & $10^{13}$ & $\mathrm{s^{-1}}$\\
$\rho$ & Regolith bulk density & $1100$ & $\mathrm{kg\,m^{-3}}$\\
$\sigma$ & Stefan--Boltzmann constant & $5.6703271\cdot 10^{-8}$ & $\mathrm{kg\,s^{-3}\,K^{-4}}$\\
\hline 
\hline
\end{tabular}
\caption{A summary of model constants and their numerical values used in this work. The numerical values for 
$A$, $c$, $P$, $\varepsilon$, $\kappa$, and $\rho$ are taken from \protect\citet{hayneetal17}. The 
lunar spin pole orientation $\{\alpha_{\rm L},\delta_{\rm L}\}$ was obtained from the formulae given by 
\protect\citet{roncoli05} and evaluated at J2000.}
\label{tab2}
\end{center}
\end{table*}

\section{The desorption model} \label{sec_desorption}

Desorption of water molecules from the lunar surface is here modelled by solving the Polanyi--Wigner equation,
\begin{equation} \label{eq:05}
\frac{df}{dt}=-\nu\exp\left(-\frac{\Delta E_{\rm des}}{k_{\rm B}T}\right)f^n,
\end{equation}
see, e.g., \citet{hibbittsetal11}; \citet{nobleetal12}; \citet{collingsetal15}; \citet{suhasariaetal17}. 
Descriptions of parameters and numerical values of constants are given in Tables~\ref{tab1} and \ref{tab2}. The Polanyi--Wigner 
equation is here solved numerically for each facet with the $4^{\rm th}$ order Runge--Kutta method \citep[e.g.][]{burdenfairs93}, using 
surface temperatures $T=T(0,t)$ obtained with the thermophysical modelling in Sec.~\ref{sec_model}.

The values for $\nu$ and $\Delta E_{\rm des}$ applicable to $\mathrm{H_2O}$ are taken from \citet{hibbittsetal11}. Due to the 
difficulty of finding appropriate lunar regolith analogue materials, there is some uncertainty in regards to the value of the applied 
activation energy, $\Delta E_{\rm des}$, which is further discussed in Sec.~\ref{sec_discussion}. The concentration of water molecules 
on the lunar surface is extremely low -- the depth of the $3.1\,\mathrm{\mu m}$ absorption band in Chandrayaan--1/M$^3$ observations 
suggest at most a monolayer \citep{pietersetal09}. Therefore, we assume first--order desorption ($n=1$) that is appropriate for such low concentrations.

\section{Results} \label{sec_results}

\begin{figure}
\begin{center}
     \scalebox{0.4}{\includegraphics{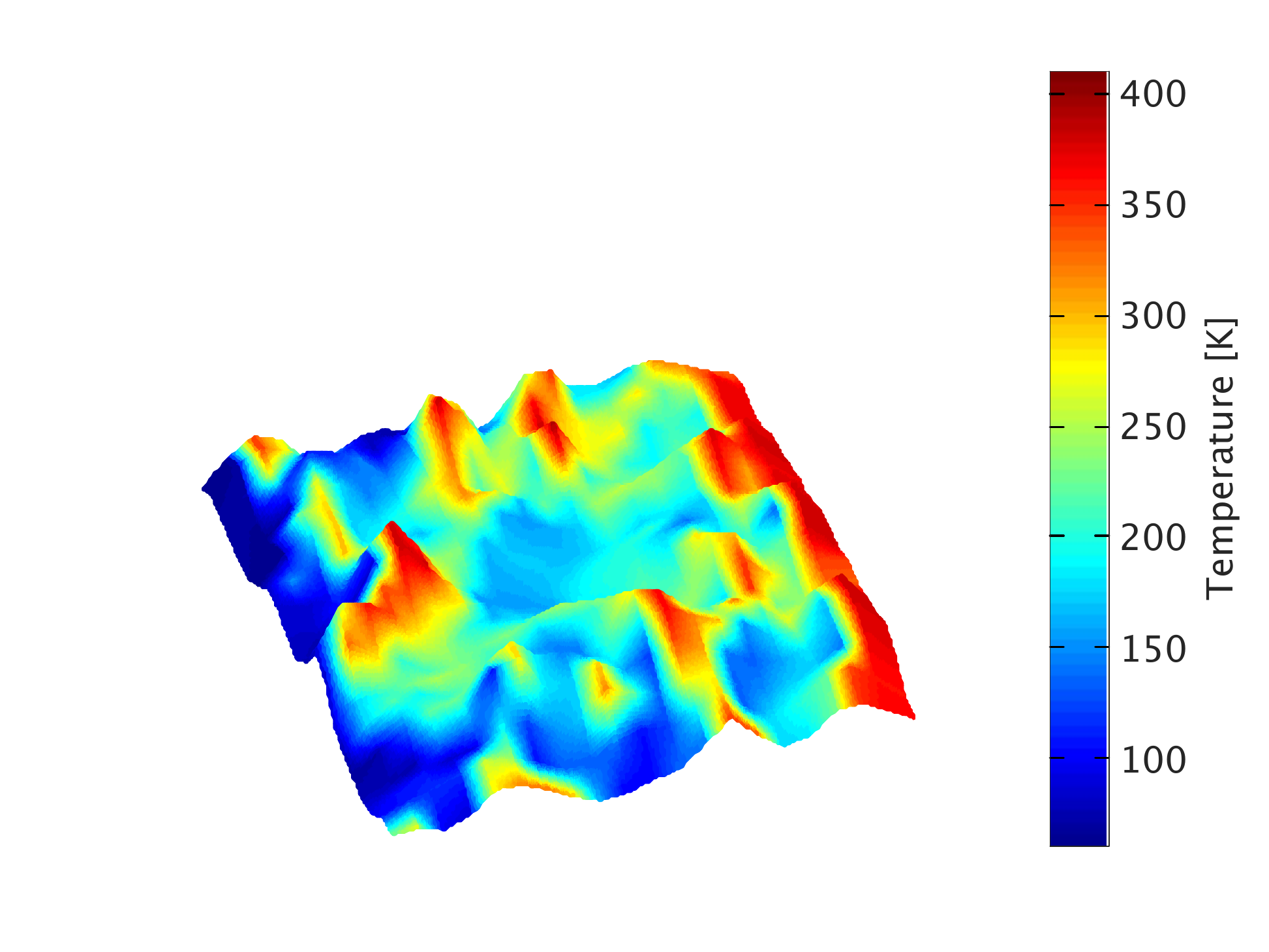}}
    \caption{This figure exemplifies the distribution of surface temperature across the rough terrain with $\xi\approx 0.4$, here 
15.77 hours after sunrise at the equator. Cold spots with temperatures down to $60\,\mathrm{K}$ are mixed with hot spots where the temperature 
approaches $390\,\mathrm{K}$. Water molecules can avoid desorption on surface patches that are shadowed by topography.}
     \label{fig_surftemp}
\end{center}
\end{figure}

\begin{figure}
\begin{center}
     \scalebox{0.4}{\includegraphics{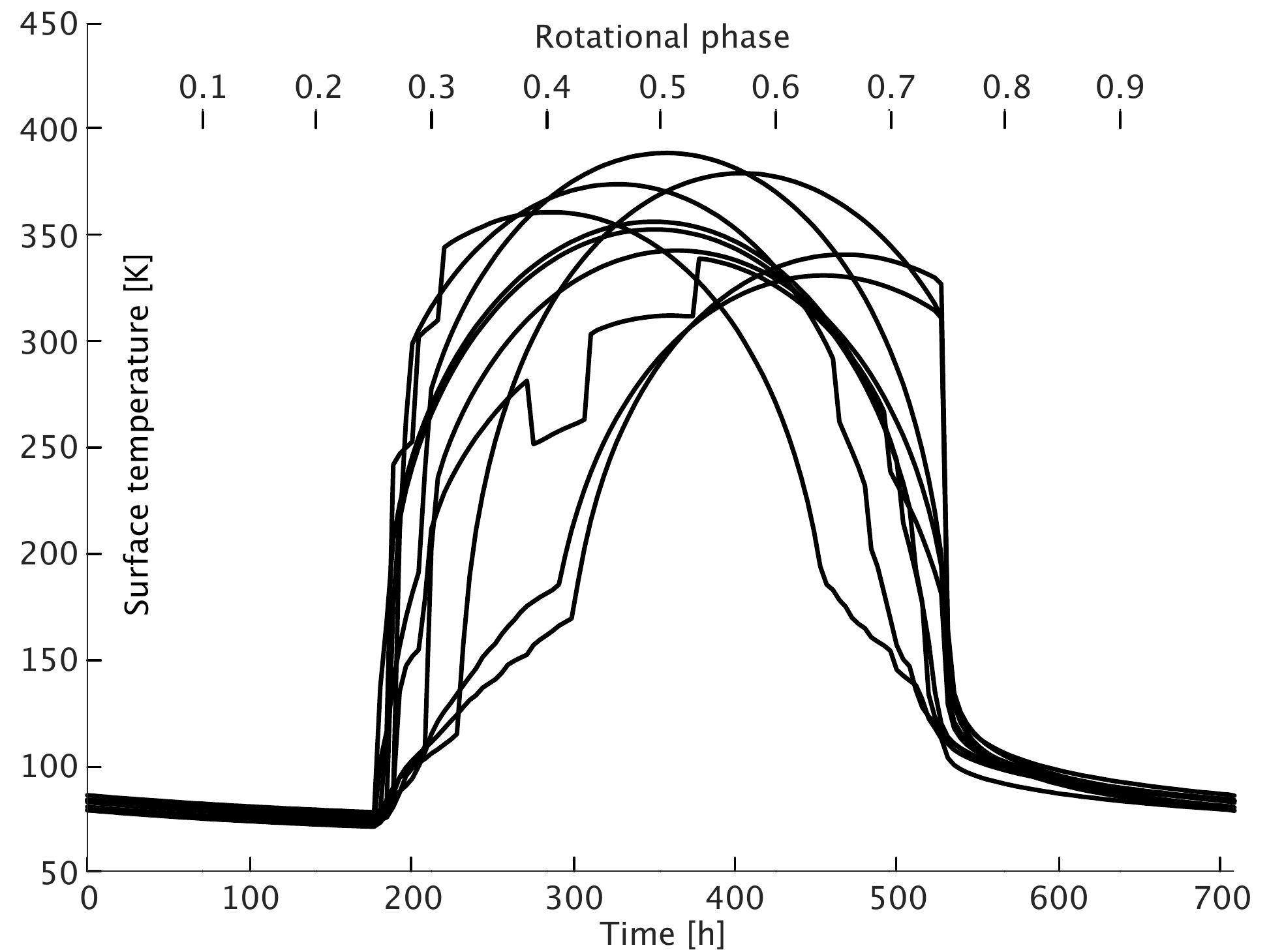}}
    \caption{Here we show the lunar surface temperature versus time (and rotational phase) during a full lunar rotation for an arbitrary selection of 
facets on a terrain with $\xi\approx 0.08$ located at $l_{\rm L}=45^{\circ}\,\mathrm{N}$ to demonstrate the effect of roughness 
on the lunar surface temperature profile.  As seen from these results, the time--evolution of temperature for a given facet depends 
on its exact orientation, and whether it is shadowed by surrounding topography during the day.}
     \label{fig_temp_time}
\end{center}
\end{figure}

In order to exemplify the effect of roughness, Fig.~\ref{fig_surftemp} shows a snapshot of the surface temperature distribution 
for the $\xi\approx 0.4$ terrain at the equator, shortly after sunrise. The temperature difference between cold-- and hot--spots exceed $300\,\mathrm{K}$ in 
this example. To further illustrate roughness effects, Fig.~\ref{fig_temp_time} shows the steady--state surface temperature as a 
function of time for a number of arbitrarily selected facets on a terrain with $\xi\approx 0.08$ located at $l_{\rm L}=45^{\circ}\,\mathrm{N}$ 
for a full lunar rotation period. Lunar surface facets that happen to be tilted toward the Sun heat very quickly and reach peak 
temperatures early in the day. However, the facets that face away from the Sun experience much slower heating (governed by the 
diffuse self--heating radiation field from surrounding illuminated facets) and experience peak heating late in the day. Examples of 
facets experiencing temporary partial shadowing are visible as well. At a given instant of time, it is therefore possible to have a 
$\sim 200\,\mathrm{K}$ variability in surface temperature within a piece of terrain that could be just $\sim 1\,\mathrm{m^2}$ or smaller. 
The desorption pattern of water molecules during sunrise is expected to be correspondingly complex.

\begin{figure}
\begin{center}
     \scalebox{0.4}{\includegraphics{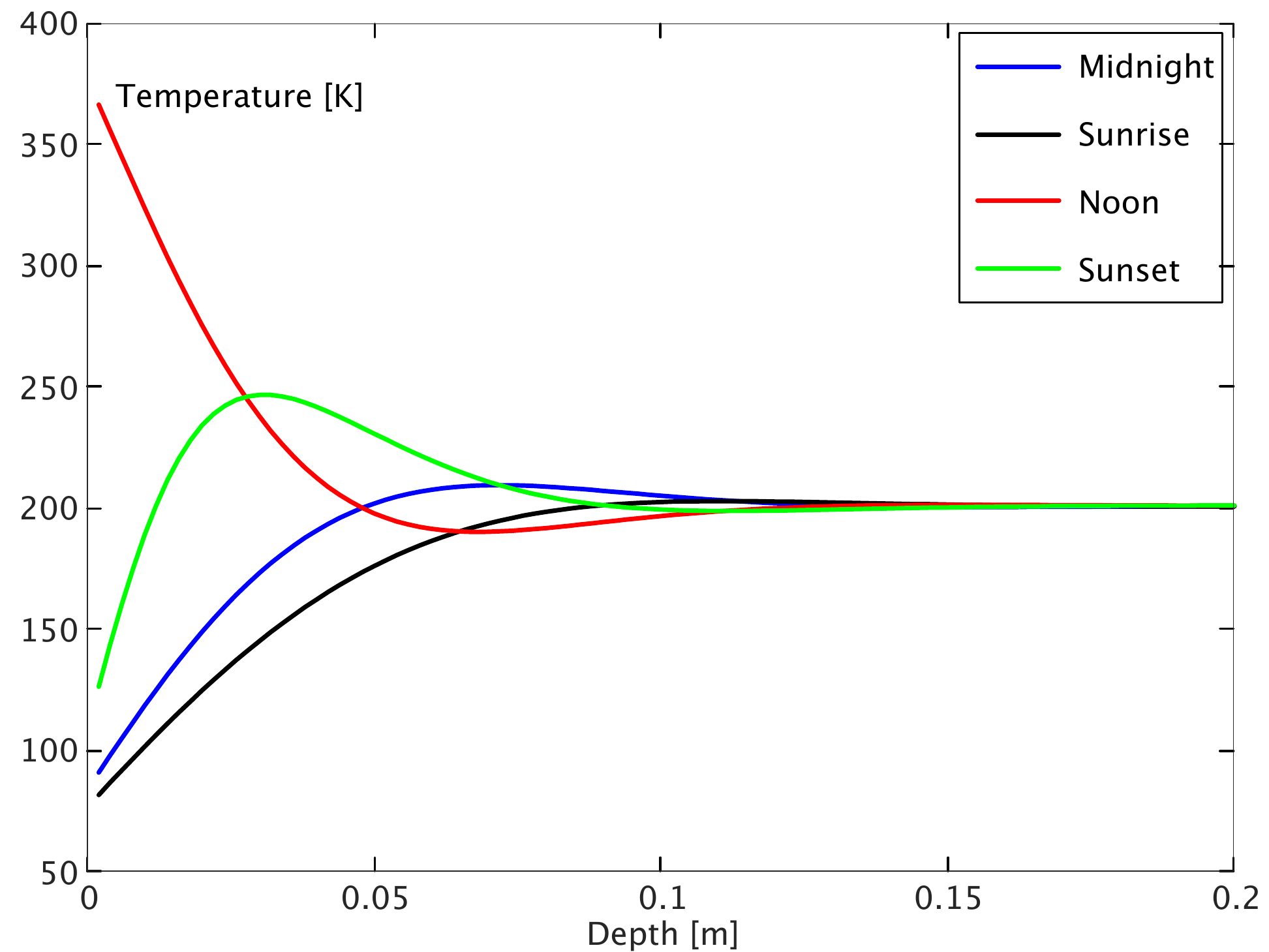}}
    \caption{Temperature as function of depth for an arbitrarily selected facet on a terrain with $\xi\approx 0.08$ located at $l_{\rm L}=45^{\circ}\,\mathrm{N}$, during four instances of the lunar day. The quasi--constant temperature of $\sim 200\,\mathrm{K}$ below 
$0.15\,\mathrm{m}$ depth shows that steady--state has been reached. }
     \label{fig_temp_depth}
\end{center}
\end{figure}

Figure~\ref{fig_temp_depth} exemplifies the temperature as a function of depth for an arbitrarily selected facet on the terrain with 
$\xi\approx 0.08$ and $l_{\rm L}=45^{\circ}\,\mathrm{N}$. The converging curves at $x\stackrel{>}{_{\sim}} 0.15\,\mathrm{m}$ with 
negligible slope show that the thermophysical solution has reached steady--state. It also serves to show how rapidly the diurnal 
temperature oscillation is damped with depth in the poorly conducting lunar regolith.

Using input from Eqs.~(\ref{eq:01})--(\ref{eq:04}) the solutions $f_j$ to the Polanyi--Wigner equation (Eq.~\ref{eq:05}) are 
used to calculate an area--weighted average of the desorbed amount of water, normalised to unity at dawn;
\begin{equation} \label{eq:06}
\mathcal{D}(t)=1-\frac{\sum_j s_jf_j(t)}{\sum_j s_j}.
\end{equation}

\begin{figure*}
\centering
\begin{tabular}{cc}
\resizebox{8cm}{8cm}{\includegraphics{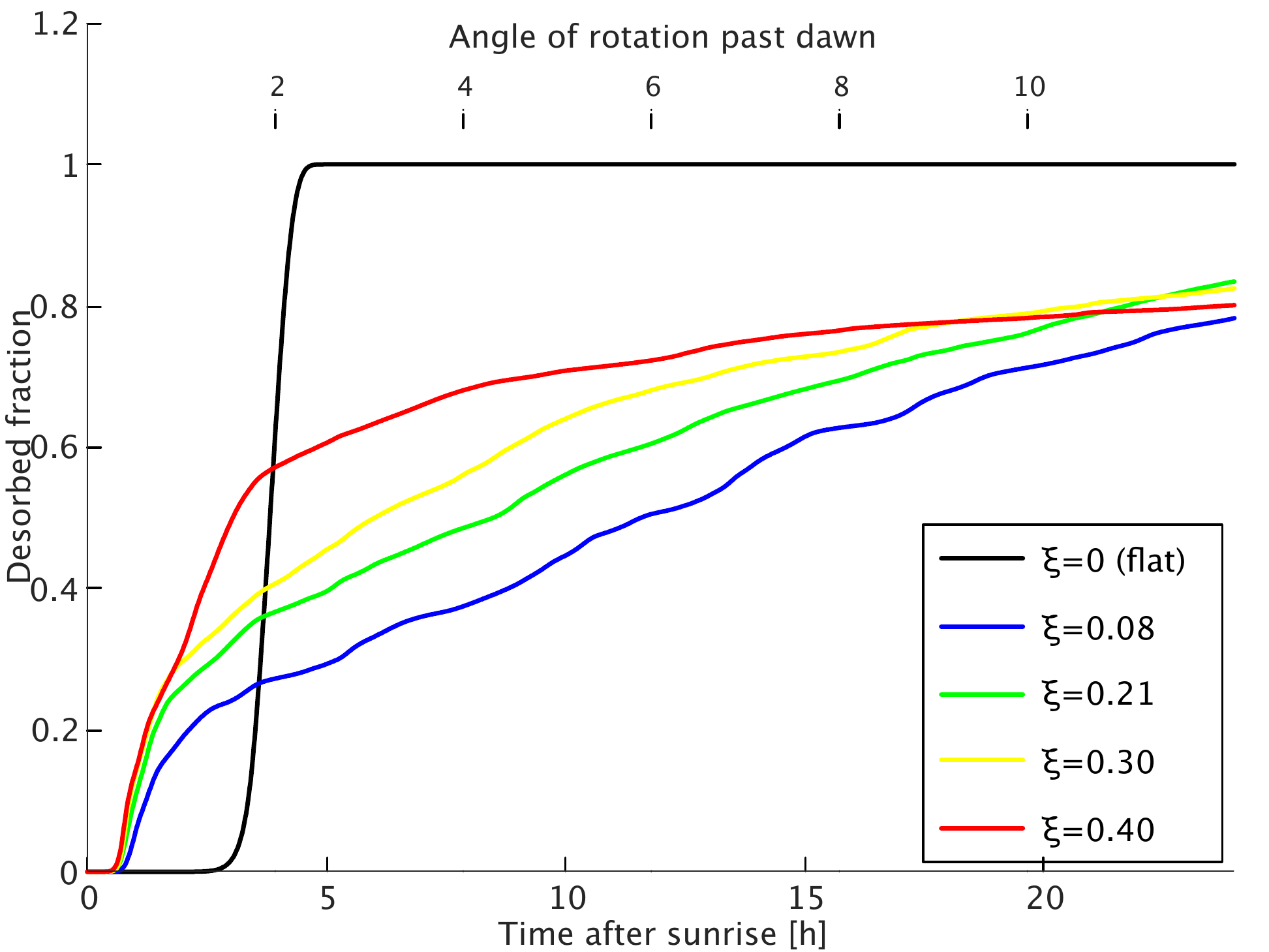}} & \resizebox{8cm}{8cm}{\includegraphics{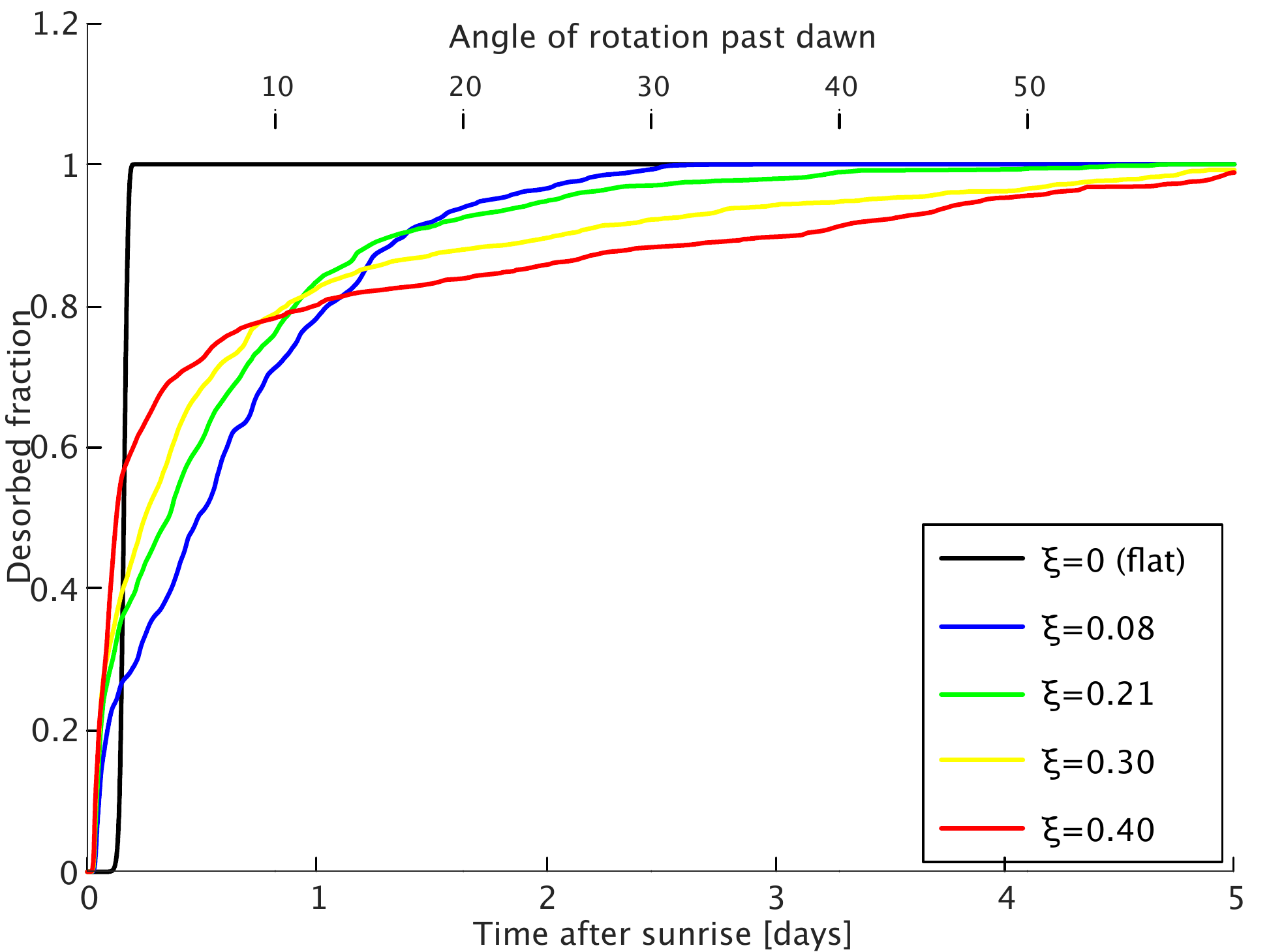}}\\
\end{tabular}
     \caption{\emph{Left:} Here we show the desorbed fraction of water $\mathcal{D}$ versus time and rotational angle past dawn at 
$l_{\rm L}=0^{\circ}$ during the first 24 h after dawn, for flat and rough terrains (different $\xi$). Although almost all of the water remains 
on the flat surface  during the first $\sim 3\,\mathrm{h}$, desorption quickly spikes after that, and almost all the water is gone by $5\,\mathrm{h}$ 
past sunrise. In contrast, for rough surfaces, the initial spike occurs earlier, at about $1\,\mathrm{h}$ after sunrise. However, the rate of desorption 
tapers off at around $\sim 2.5\,\mathrm{h}$, and the slower desorption rate means that water remains on the surface much longer than for flat terrains. 
\emph{Right:} The desorbed fraction of water $\mathcal{D}$ versus time at $l_{\rm L}=0^{\circ}$ during the first $5\,\mathrm{days}$ after dawn, for 
flat and rough terrains, extending the timeline with respect to the left panel. The rougher is the surface, the longer does it hold 
on to a fraction of its water.}
     \label{fig_eq}
\end{figure*}

Figure~\ref{fig_eq} (left) shows $\mathcal{D}(t)$ for the four rough terrains if located at the equator ($l_{\rm L}=0^{\circ}$) during the 
first $24\,\mathrm{h}$. For comparison, it also shows the results we obtain for a surface element that is flat. Water on a flat surface 
manages to hold on to the lunar regolith for the first $\sim 3\,\mathrm{h}$ but once desorption starts the water disappears into 
the exosphere very rapidly. After $\sim 5\,\mathrm{h}$, corresponding to a rotational angle of merely $2.5^{\circ}$ past local dawn, 
all water has been desorbed.

The rough surfaces display water desorption patterns that are quite different from that of a flat surface. In all cases, the rough 
surface model shows a rapid desorption in the first few hours that removes 20--50\% of the available water more quickly than for 
the flat surface. The rougher the surface, the more significant is the  fraction of water that is being removed. This initial boost 
of desorption is caused by the parts of the surface that face the Sun more directly when they first rotate across the terminator. 
Consequently, there is a period when the highest concentration of water molecules is seen on surfaces with intermediate roughness, 
while flat and highly rough terrains are both dryer. However, the rate by which $\mathcal{D}$ changes then quickly decreases, 
and substantial fractions of the water molecules linger on the surface, even when the degree of roughness is rather mild. The curves 
converge, and by $t=24\,\mathrm{h}$ about $\sim 20\%$ of the originally condensed $\mathrm{H_2O}$ molecules are still located on the regolith, regardless of degree of roughness as long as $\xi\not=0$.

Figure~\ref{fig_eq} (right) shows the results of the same simulations as in its left panel but extended to the first five days after lunar dawn. 
Here, an inversion point around $t\approx 1\,\mathrm{days}$ is seen where the high--$\xi$ $\mathcal{D}$ curves for the first time dive 
under the low--$\xi$ curves. The $\xi\approx 0.08$ surface reaches $\mathcal{D}=0$ after $t\approx 2.5\,\mathrm{days}$ when $1/3$ of the rotational angular distance to noon has been crossed. 
At that time the $\xi\geq 0.3$ surfaces still hold on to more than 10\% of their original water molecule coverage. Those molecules are hiding on the hill 
sides that slope away from the Sun that are relatively cool. Small amounts of water is desorbing into the exosphere even five days after sunrise.

\begin{figure*}
\centering
\begin{tabular}{cc}
\resizebox{8cm}{8cm}{\includegraphics{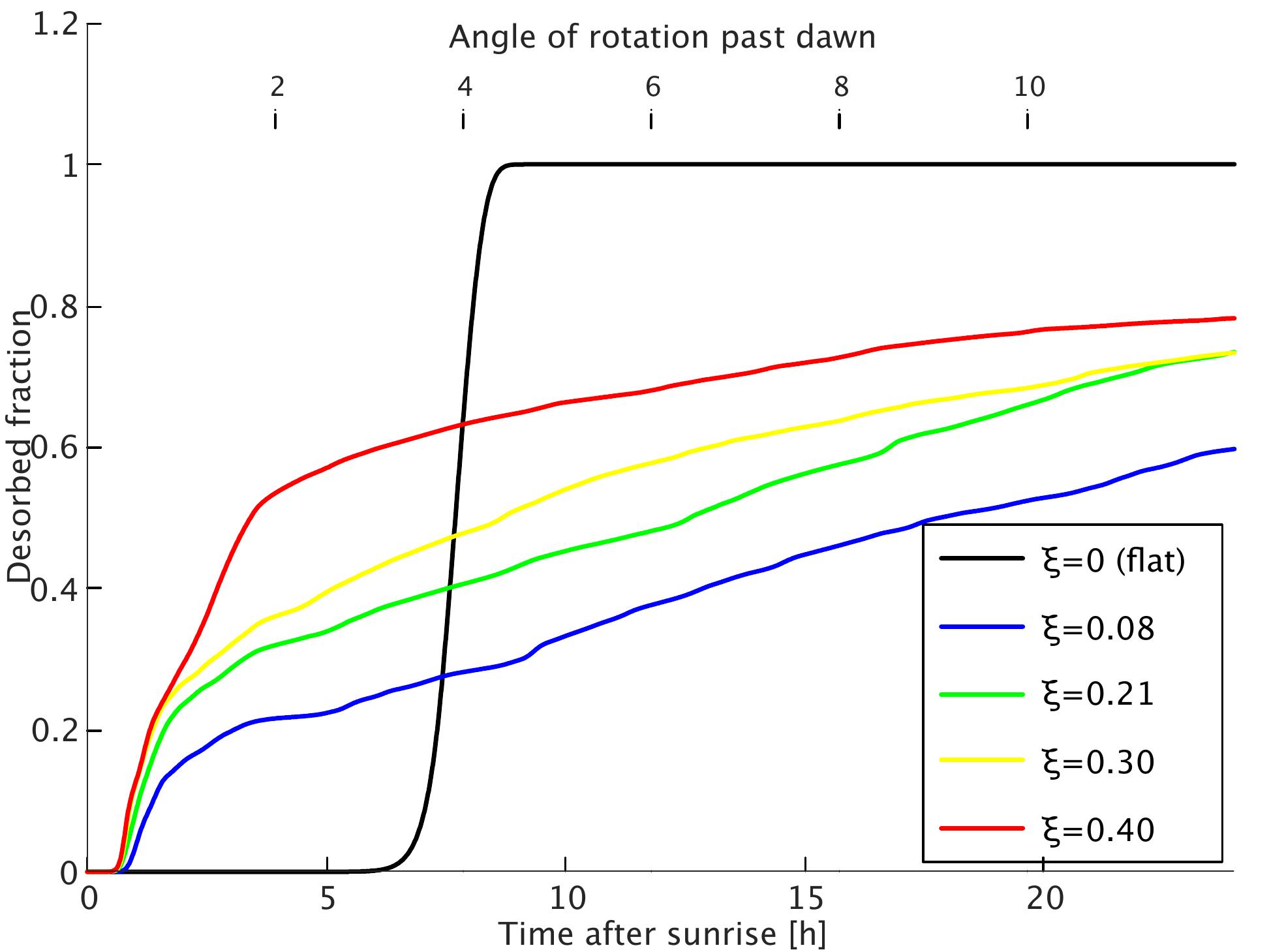}} & \resizebox{8cm}{8cm}{\includegraphics{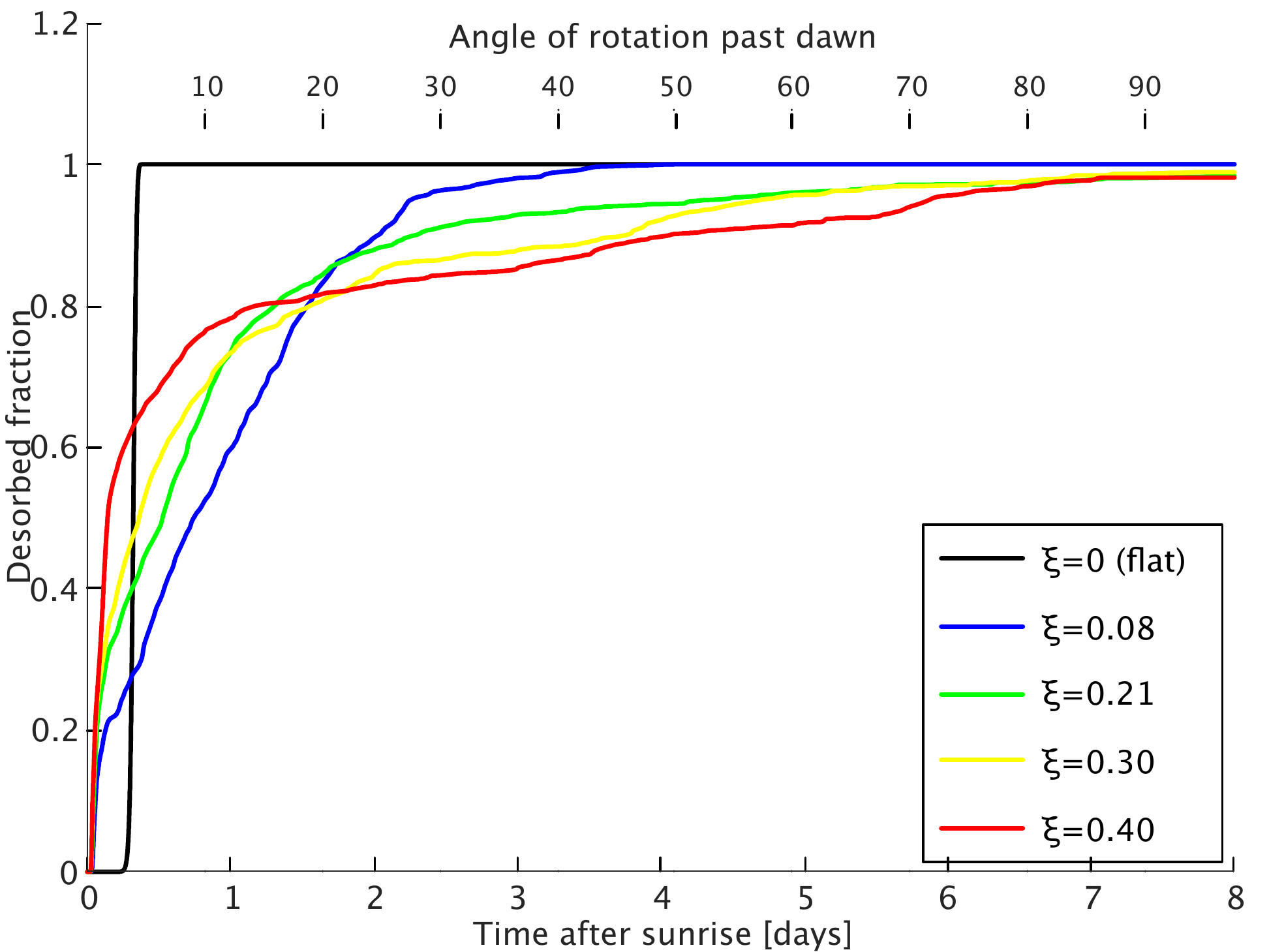}}\\
\end{tabular}
     \caption{\emph{Left:} The desorbed fraction of water  $\mathcal{D}$ versus time and rotational angle past dawn at 
$l_{\rm L}=45^{\circ}\,\mathrm{N}$ during the first $24\,\mathrm{h}$ after dawn, for flat and rough terrains. Almost all of the water 
remains on the flat surface  during the first  $\sim 6\,\mathrm{h}$, then desorption quickly spikes after that, and almost all the water 
is gone by $8\,\mathrm{h}$ past sunrise. The time--scales for water loss from flat surfaces are similar at latitudes $0^{\circ}$ and $45^{\circ}$.  
Rough surfaces lose water rapidly within the first hour after sunrise. However, the rate of desorption tapers off at around $\sim 2.5\,\mathrm{h}$, 
and the slower desorption rate means that rough surfaces retain a fraction of the water substantially longer than a flat surface. It is worth noting here 
that the roughest surface is very similar to the case of $l_{\rm L}=0^{\circ}$ (see Fig.~\ref{fig_eq}). In contrast, the completely flat 
surface is significantly affected by the change in lunar latitude. \emph{Right:} The desorbed fraction of water $\mathcal{D}$ versus time 
at $l_{\rm L}=45^{\circ}\,\mathrm{N}$ during the first $8\,\mathrm{days}$ after dawn, for flat and rough terrains. Similarly to the case of 
$l_{\rm L}=0^{\circ}$ in Fig.~\ref{fig_eq}, the extended timeline shows that the rougher the surface, the more water remains on the lunar 
surface over time. The roughest surfaces at this latitude are capable of preserving a small fraction of the water until local noon, which was 
not the case at the equator.}
     \label{fig_lat45}
\end{figure*}

At higher lunar latitudes the intensity of solar illumination is weaker than at the equator and the desorption is expected to be slower. 
Figure~\ref{fig_lat45} (left) is a close--up on $\mathcal{D}(t)$ evolution during the first $24\,\mathrm{h}$ at $l_{\rm L}=45^{\circ}\,\mathrm{N}$. 
Here, water on flat surfaces is gone within $\sim 8\,\mathrm{h}$ after dawn while modestly rough terrain still holds on to $\sim 70\%$ of the 
initial amount of water at that time. Because of the relatively high desorption rate of very rough surfaces prior to the inversion point, the 
$\xi\geq 0.3$ surfaces retain 35--50\% of the water $8\,\mathrm{h}$ after dawn.

Figure~\ref{fig_lat45} (right) shows the same simulations as its left panel, except that the time axis extends to eight days after sunrise, i.e., 
somewhat beyond noon. The inversion point takes place $\sim 1.5\,\mathrm{days}$ after dawn, and the $\xi\approx 0.08$ terrain becomes dry about 
$\sim 3.5\,\mathrm{days}$ after sunrise, which is a day longer than at the equator. At higher degrees of roughness ($\xi\stackrel{>}{_{\sim}} 0.2$), 
a few percent of the water manages to remain attached to the lunar surface up to and beyond noon. At these latitudes, there are some 
surface facets on relatively steep slopes that never experience direct sunlight, and the diffuse radiation flux from self-heating is not 
sufficiently strong to desorb all water. That suggests the presence of small permanent cold traps sprinkled across the lunar dayside at mid--to--high latitudes. 

\begin{figure*}
\centering
\begin{tabular}{cc}
\resizebox{8cm}{8cm}{\includegraphics{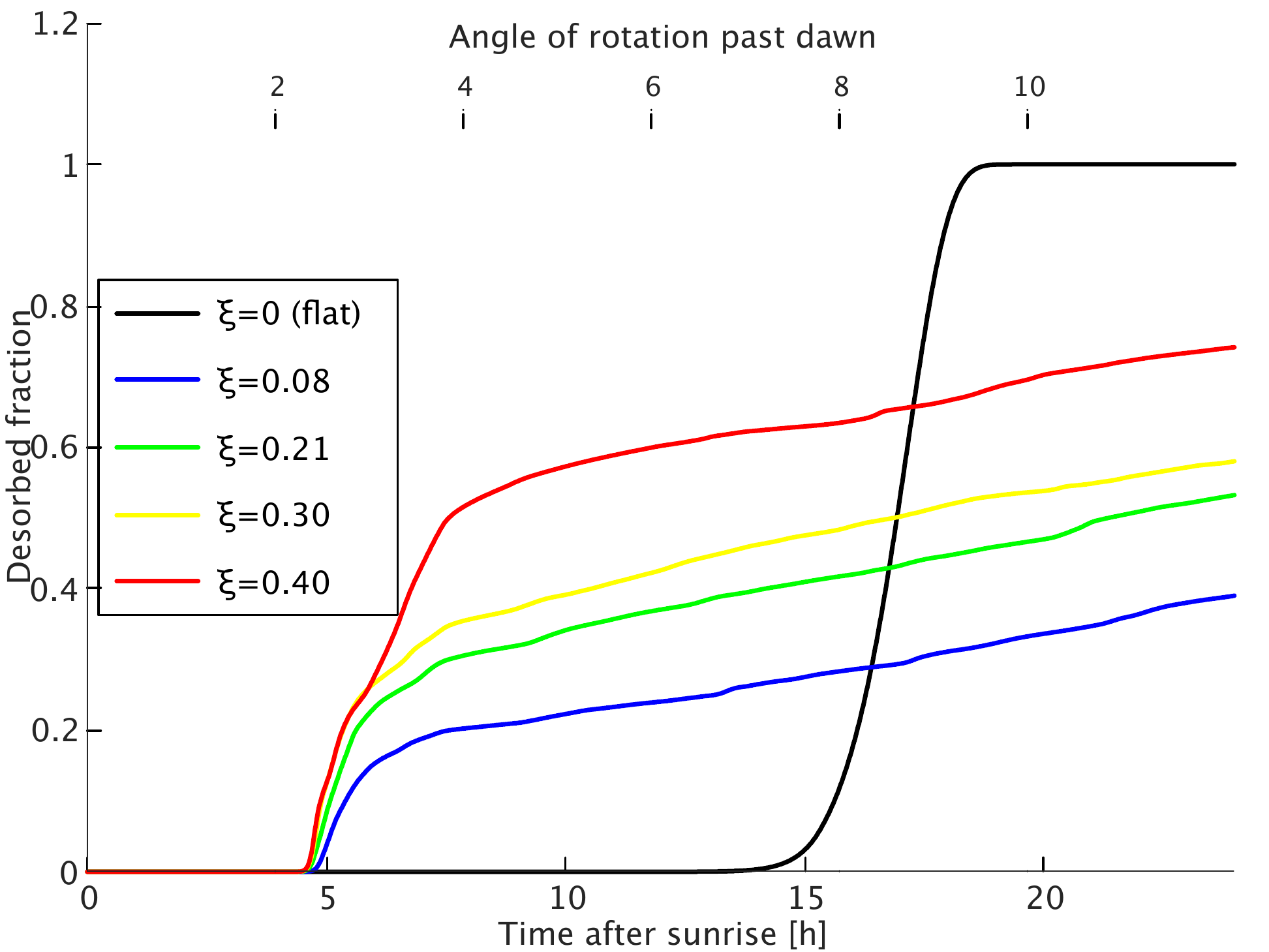}} & \resizebox{8cm}{8cm}{\includegraphics{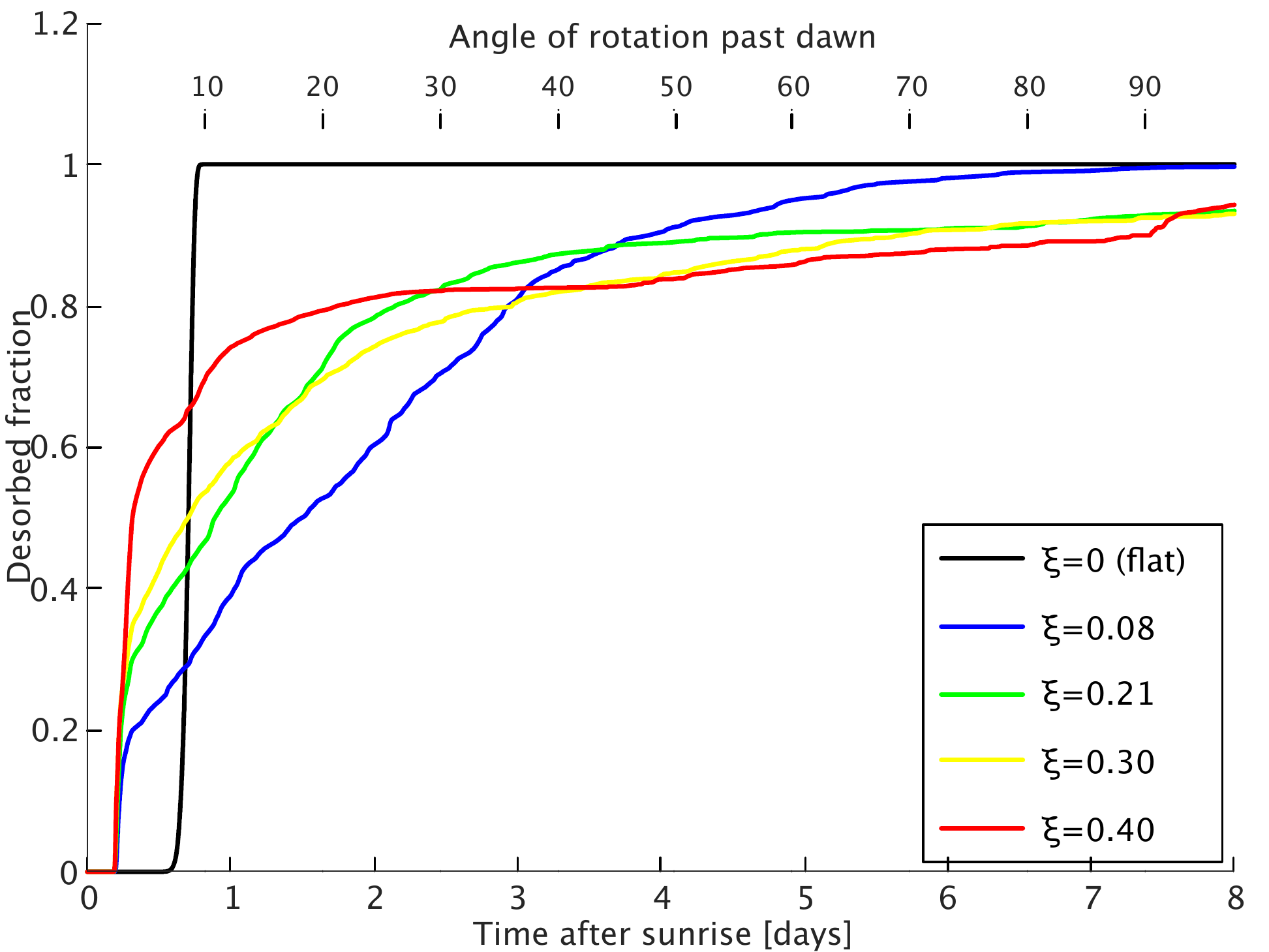}}\\
\end{tabular}
     \caption{\emph{Left:} The desorbed fraction of water  $\mathcal{D}$ versus time and rotational angle past dawn at 
$l_{\rm L}=70^{\circ}\,\mathrm{N}$ during the first $24\,\mathrm{h}$ after dawn, for flat and rough terrains. Compared to equatorial to 
mid--latitude locations, the desorption is delayed quite substantially for flat terrain, yet all water is lost within $20\,\mathrm{h}$. \emph{Right:} The desorbed 
fraction of water $\mathcal{D}$ versus time at $l_{\rm L}=70^{\circ}\,\mathrm{N}$ during the first $8\,\mathrm{days}$ after dawn, for flat and rough terrains. The 
tendency for rough surfaces to hold on to a fraction of the water all the way up to noon is even stronger than at $l_{\rm L}=45^{\circ}$.}
     \label{fig_lat70}
\end{figure*}

These effects are even stronger at  $l_{\rm L}=70^{\circ}\,\mathrm{N}$, shown in Fig.~\ref{fig_lat70}. Because of the lower solar 
fluxes at high latitudes, water remain on a locally flat surface during the first $15\,\mathrm{h}$ after dawn, but has desorbed completely within one Earth--day. 
Meanwhile, the rough surfaces lose $30\%$--$70\%$ of the water depending on the $\xi$--value, but at that point the loss rate rapidly decreases. The $\xi\geq 0.21$ 
terrains hold on to about $10\%$ to their water all the way up to local noon. Near the pole (at $l_{\rm L}=85^{\circ}\,\mathrm{N}$), that number increases slightly, to $17\%$.

Surface roughness has an essential effect on the water desorption rate, resulting in a stronger release of water into the exosphere 
near the dawn terminator and an enhanced capacity of lunar regolith to retain water at later times compared to the idealised flat terrain. 
These effects are prominent at the equator and grow increasingly stronger with latitude. The structure and properties of the modelled lunar 
exosphere are presumably highly sensitive to assumptions made about the lunar surface roughness and its effect on local temperature. 
With our  improved model, we predict the presence of water permanently existing in cold traps located at mid--to--high latitudes, 
thus challenging the notion \citep[e.g.][]{clark09, honniballetal20} that water would not be able to exist on the 
lunar dayside as an adsorbed volatile.

\section{Discussion} \label{sec_discussion}

The lunar exosphere and its interaction with the surface of the Moon embody several highly complex phenomena because pre--existing 
water molecules trapped on the cold night--side surface regolith can feed the exosphere when carried into sunlight by lunar rotation. 
In addition, water molecules are formed continuously on the sunlit lunar hemisphere where the regolith interacts with the solar wind 
\citep[e.g.][]{ichimuraetal12}, and creates additional mass source terms for the exosphere.  Numerous processes remove newly 
formed or pre--existing molecules from the surface besides thermal desorption, such as photon, charged--particle, and chemical 
sputtering \citep{stern99}, all associated with different exosphere injection rates with or without temperature dependence. 
Thus, we do not claim to have performed exhaustive modelling of the surface--exosphere interaction phenomena but are using a subset 
of processes (thermal desorption of pre--existing water molecules) to illustrate and quantify the effects of surface roughness and 
(sub)monolayer desorption on the dayside lunar water surface coverage. We find that surface roughness simultaneously greatly 
enhances the desorption rate of water near the dawn terminator, \emph{and} the capability of the surface to hold on to a fraction of the 
water late in the day (in some cases up to and beyond noon), compared to an idealised flat surface. Keeping in mind that the facet 
orientations that protect molecules from strong solar illumination also shield them from direct solar wind exposure, roughness will also 
have a significant effect on physical processes other than thermal desorption.

Though non--exhaustive, the simulations presented in this paper have important implications for the interpretation of observations of 
lunar dayside surface water. For instance, \citet{clark09} dismissed the possibility that the water molecules observed by Cassini 
were simply condensed onto the lunar regolith based on the perceived lack of stability of ice at the average temperatures found 
on the lunar dayside and suggested that the observed water was trapped within the mineralogical structure of the regolith grains. 
A similar statement was put forward by \citet{honniballetal20}, who argued that the majority of the observed  
molecular $\mathrm{H_2O}$ was trapped within impact glass. The main reason for that assumption was the perceived difficulty of 
preserving desorbed water on the lunar dayside. If one assumes that the lunar surface is an idealised flat terrain (as in the $\xi=0$ 
solid black curves in Figs.~\ref{fig_eq}--\ref{fig_lat45}), the \citet{clark09} and \citet{honniballetal20} conclusions can certainly 
be justified. In fact, the latter paper explicitly quoted thermophysical models that assumed a locally flat lunar surface. 
However, when modelling the effects of surface roughness on the desorption rate, it becomes clear that the Moon, continuously and at 
any local time of day, exposes surface elements to the Sun and to an observing spacecraft that has received little previous heating and 
retain volatiles. It is therefore not surprising that water molecule absorption features may be observed on large parts of the lunar dayside, 
and that the absorption line depth shows a significant anticorrelation with local illumination levels, as observed on the 
Moon \citep{sunshineetal09}, which is inconsistent with the sole presence of hydrated minerals. The possibility of adsorbed water existing 
in select locations on the  entire lunar dayside opens up a completely new avenue of thought in the analysis of current and future 
measurements of lunar water molecules. Furthermore, the accuracy of exospheric models will be substantially improved if the source 
and sink terms of exospheric water includes a description of the lunar surface that accounts for its roughness.

During the final preparation of this paper, \citet{hayneetal20} published an investigation that also stresses 
the importance of surface roughness in the context of understanding presence of adsorbed water on the lunar dayside. They presented 
thermophysical simulations and temperatures of the rough lunar surface using a code similar to ours. \citet{hayneetal20} 
conclude that shadowed regions equatorward of latitude $80^{\circ}$ are too warm to support accumulation of water, although 
they did not consider the actual desorption problem in detail. Based on our work, we are more optimistic about presence of 
adsorbed water on the lunar surface, even at noon and rather close to the equator. We acknowledge, however, that the applied activation 
energy $\Delta E_{\rm des}$ used in Eq.~(\ref{eq:05}) is an important source of uncertainty in models, as mentioned in Sec.~\ref{sec_desorption}. 
Such material parameters are measured routinely in the laboratory for different combinations of substrate materials and adsorbed volatiles 
\citep[e.g.][]{nobleetal12, kerisitetal13, collingsetal15, suhasariaetal17, mclainetal18}. In these particular simulations, we relied on the 
values provided by \citet{hibbittsetal11}. The difficulty consists of finding appropriate analogue materials that accurately represent 
lunar regolith. Even lunar samples may not be sufficiently representative, as illustrated by the work of \citet{bernatowicsandpodosek91}. 
They measured the adsorption of argon on crushed lunar rocks and noted rapid changes of the measured capacity of the rock to hold on to 
the volatile over time. They attributed the initially high measured activation energies to a particular mineralogical structure of freshly exposed 
rock surfaces formed through radiation damage and hydrogen--saturation in the lunar environment and interpreted the rapid deterioration of 
these properties as due to interaction between the rock and the terrestrial atmosphere. If there indeed are significant differences between 
the capability of real lunar regolith to bind volatiles, to those of terrestrial rocks or lunar samples brought to Earth, we may substantially 
overestimate the desorption rate of volatiles in the lunar environment in current models.

We consider this problem a necessary reason to intensify \emph{in situ} lunar measurements in order to have a better understanding of 
the behaviour of lunar regolith in its natural environment. For example, if the number density of exospheric water and its temporal evolution 
during lunar rotation could be measured accurately from various altitudes in various times of a lunar day, and if theoretical models are 
evolved to a point where they accurately account for the most important physical processes that govern surface--exosphere interaction, 
then material parameters such as $\Delta E_{\rm des}$ could in principle be retrieved by forcing a match between calculations and observations. 
This could provide insight into the physical properties of the regolith that are difficult or impossible to establish in terrestrial laboratories.

Future exospheric models of ballistically jumping molecules should include a realistic description of surface roughness. For a given 
location and local time on the lunar surface, a thermophysical model similar to the one in the current paper should be used to calculate 
a realistic temperature distribution. At each molecular interaction with the surface, a temperature should be drawn from that distribution 
to simulate the presence of local hot--spots and cold--spots. This will diversify the range of local residence times and ejection velocities 
compared to flat--surface models. As a consequence, the steady--state number density distribution of lunar exosphere models would 
adjust accordingly, and become more realistic. Such improvements will prove valuable, once the observational technology has advanced 
to the point where the lunar water number density can be measured accurately as function of latitude, altitude, and local time. 

If the current lunar exploration trend continues, which inevitably brings terrestrial water and other substances artificially to the Moon, 
there is a danger of losing the opportunity to understand the natural lunar environment. Consequently, it is of critical importance and urgency 
to determine the physical properties (e.g., density and temperature), composition, and time variability of the water and other volatiles in the 
fragile lunar exosphere, before being perturbed by further human activity. Scientific testable hypotheses drive the development of 
instrumentation that fly on spacecraft, as well as the definition of goals and objectives that motive the measurements and experiments proposed for 
(and carried out during) missions. It is therefore urgently important to continue the development of realistic models of the lunar surface and 
exosphere, and their complex interactions. Results from such models need to be accurate and relevant, as well as capable of generating 
predictions that connect observable phenomena with fundamental questions about the formation and evolution of the Earth--Moon system. 
This is the prerequisite for formulating science goals considered sufficiently compelling to motivate mission selection. We consider this paper 
as being part of such a community effort, and hope that our work on prolonged survival of adsorbed water in temporary cold--spots on 
the lunar dayside will inspire to the development of instruments and experiments suitable to study this phenomenon \emph{in situ}.

\section{Conclusions} \label{sec_conclusions}

The model of the surface temperature of the Moon described in this paper has significant implications for understanding the 
presence and evolution of water on the lunar surface. Previously considered models often discounted the roughness of the lunar surface. 
As higher temperatures will increase the desorption rate of water, the main factor affecting the amount of water present on the lunar 
surface is the surface temperature. This report illustrates that the roughness of real terrains on the lunar surface creates shadowed cold spots 
and self--heated hot spots, which can cause temperature fluctuations of $\Delta T\leq 200\,\mathrm{K}$ on sub--metre level. Therefore, 
it is of critical importance to take account of the surface roughness to get an accurate picture of the amount of water on the surface of the Moon, 
and imprecise conclusions about the presence of water on the lunar surface are concluded from the idealised flat terrain assumption. 
Our thermophysical model \citep{davidssonandrickman14}, accounts for direct solar illumination, shadows cast by topography, 
visual and infrared illumination emanating from other parts of the surface, and heat conduction, without relying on arbitrary assumptions 
that violate energy conservation. The model presented here addresses molecular residence time in a manner that is closer to the 
realistic conditions than in previous models, by modelling the desorption of water molecules from the lunar surface by solving the 
Polanyi--Wigner equation. Considering lunar rough terrains have significant implications for understanding the water molecule cycle 
in the lunar environment, particularly for analysing observed properties of the exosphere, and to model the interaction between exospheric 
water with the lunar surface. The model here shows a sufficiently high degree of surface roughness substantially extends the life--time 
of adsorbed water molecules on the lunar dayside. At the equator, smooth surfaces lose all water within $\sim 3\,\mathrm{h}$ after sunrise, 
but rough terrains maintain a fraction of the water for up to 5 days. Within $45^{\circ}$ of the poles, water is still present at local noon.  
At latitude $70^{\circ}\,\mathrm{N}$, about $10\%$ of the adsorbed water at sunrise is still present at noon, increasing to $\sim 17\%$ at $85^{\circ}\,\mathrm{N}$.
These results show the surface roughness strongly controls how the lunar surface acts as a source and sink of water. 
Therefore, models of the exosphere need to account for surface roughness in order to be accurate. This is important in the context of providing modelling 
support during instrument development, mission definition, and operations, that are prerequisites for \emph{in situ} lunar explorations.

\section*{Acknowledgements} 

This research was carried out at the Jet Propulsion Laboratory, California Institute of Technology, under 
a contract with the National Aeronautics and Space Administration. Part of this work was funded by the 
Space Technology Office at JPL. We gratefully acknowledge the assistance provided by Andrew Shapiro, Kalind Carpenter, 
Kiran Hemkins, and Pamela Clarke, and thank for their immeasurable support and valuable discussions on the issues and progress of this research.\\

\noindent
\emph{COPYRIGHT}.  \textcopyright\,2021. California Institute of Technology. Government sponsorship acknowledged.

\section*{Data Availability}

The data underlying this article will be shared on reasonable request to the corresponding author.

\bibliography{MN-21-0499-MJ.R2.bbl}

\bsp	
\label{lastpage}
\end{document}